\documentclass[pdflatex,sn-mathphys-num]{sn-jnl}% Math and Physical Sciences Numbered Reference Style
\usepackage{graphicx}%
\usepackage{multirow}%
\usepackage{amsmath,amssymb,amsfonts}%
\usepackage{amsthm}%
\usepackage{mathrsfs}%
\usepackage[title]{appendix}%
\usepackage{xcolor}%
\usepackage{textcomp}%
\usepackage{manyfoot}%
\usepackage{booktabs}%
\usepackage{algorithm}%
\usepackage{algorithmicx}%
\usepackage{algpseudocode}%
\usepackage{listings}%
\theoremstyle{thmstyleone}%
\theoremstyle{thmstyletwo}%

\theoremstyle{thmstylethree}%

\begin{document}

\title[Article Title]{Performance and Stability Studies of GEM Detectors for High Energy Physics Applications}


\author*[1]{\fnm{S.} \sur{Mandal}}\email{subirmandal50@gmail.com (S. Mandal)}

\author[1]{\fnm{S.} \sur{Das}}\email{supriya@jcbose.ac.in (S. Das)}

\author[1]{\fnm{S.} \sur{Biswas}}\email{saikat@jcbose.ac.in (S. Biswas)}

\affil[1]{\orgdiv{Department of Physical Sciences},
\orgname{Bose Institute},
\orgaddress{\street{EN-80, Sector V, Bidhan Nagar},
\city{Kolkata},
\postcode{700091},
\country{India}}}


\abstract{{R\&D on different gas-filled detectors such as Gas Electron Multipliers (GEMs), Resistive Plate Chambers (RPCs) and straw tubes has been carried out in the detector laboratory of Bose Institute for more than a decade, keeping in mind present and future high-energy physics experiments. In this article, performance studies of a single-mask triple-GEM detector using a $^{55}$Fe X-ray source with an Ar/CO$_2$ (70/30) gas mixture are described. The study focuses mainly on the gain, energy resolution, count rate, and stability of the chamber under long-term irradiation. The experimental setup, methodology, and analysis results are presented.}}




\maketitle

\section{Introduction}
In the Experimental High Energy Physics (EHEP) detector laboratory of Bose Institute, Kolkata, the R\&D of different gas filled detectors such as Gas Electron Multiplier (GEM), Resistive Plate Chamber (RPC) and straw tube detector is ongoing for future heavy ion physics experiments, keeping in mind high particle rate handling capacity. The main goal of the research program is to build gaseous detectors using indigenous materials and new technology along with the characterisation and particularly the stability study of those gaseous detectors. The latest information of the performance studies of the straw tube and RPC are discussed in Reference~\cite{straw,rpc}. In this particular article some aspects of characteristic studies of GEM detector is presented.

\begin{figure}[htb]
\centering
\includegraphics[scale=0.4]{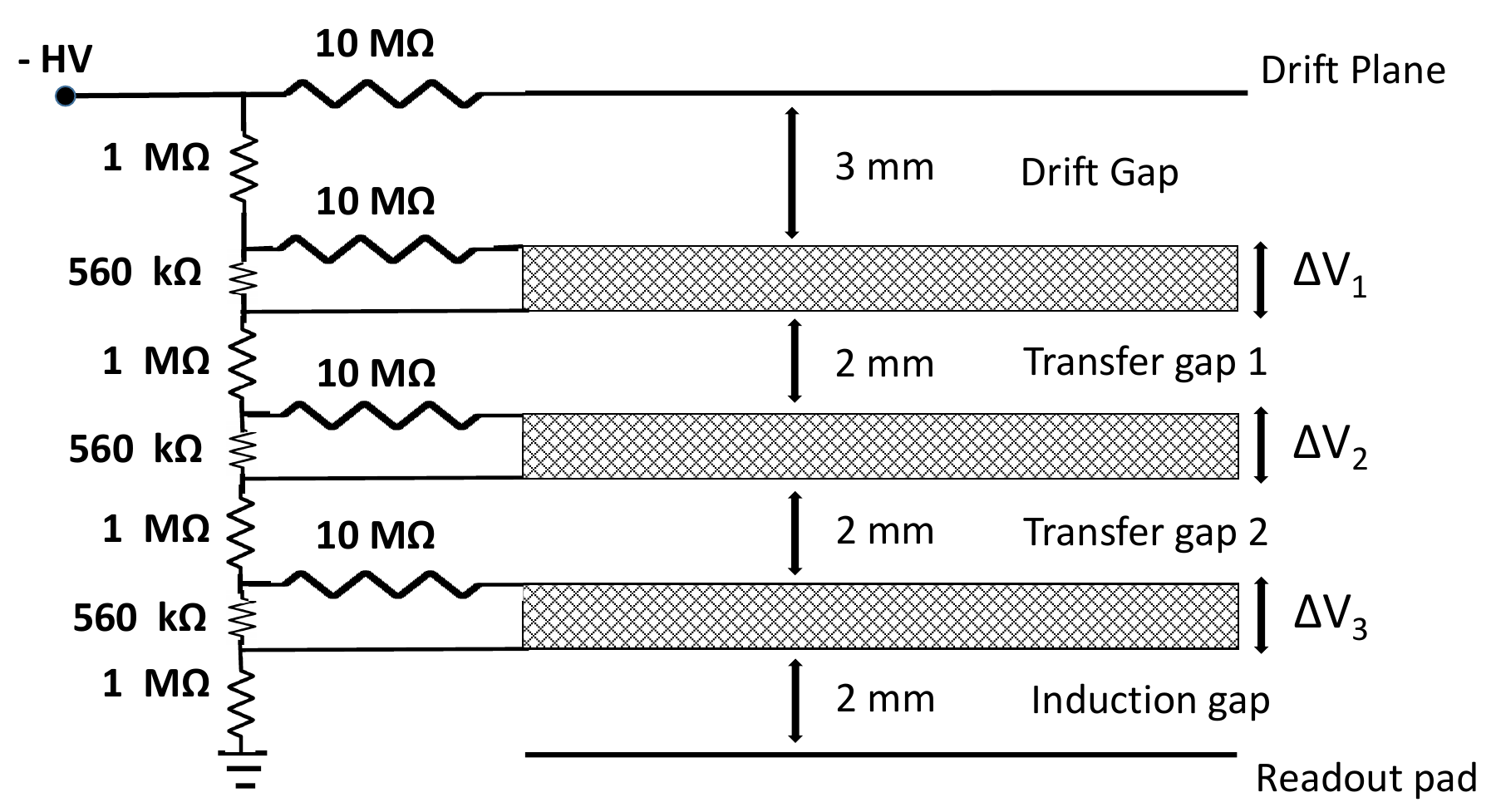}
\caption{\label{fig1} Schematic of the triple-GEM detector and the high voltage (HV) distribution network.}
\end{figure}

GEM detectors introduced by Fabio Sauli in 1997~\cite{sauli}, have become one of the most widely used gaseous detectors in high energy physics experiments~\cite{buzulutskov,ketzer,biswas}. Popularity arises because of their high rate handling capability, good spatial resolution, and stable operation under high radiation. GEM detectors belong to the class of Micro-Pattern Gaseous Detectors (MPGDs). A GEM foil is made of a thin (50~$\mu$m thickness) polyimide foil (usually Kapton) and copper cladding (5~$\mu$m thickness) on both sides~\cite{Gem_sauli_application,pinto,karadzhinova}. Photolithographic techniques are being used to create the holes inside the copper cladded Kapton foil~\cite{GEM_foil,rd51,chatterjee_review}. These microscopic holes are made with high density (typically 50-100~mm$^{-2}$) having a typical diameter 70~$\mu$m and pitch 140~$\mu$m. The GEM foils are classified as either Double Mask~(DM) or Single Mask~(SM) types, depending on the photolithographic technique used at the time of the production. A triple-GEM detector consists of a drift region followed by three GEM foils separated by transfer gaps and an induction region before the readout plane as shown in Fig~\ref{fig1}. Primary electrons produced by ionising radiation in the drift region are transported into the GEM holes, where the strong electric field produces avalanche multiplication. Successive multiplication in the three GEM foils provides sufficient charge amplification for signal detection at the readout plane~\cite{ketzer,bachmann,shah}. Triple-GEM detectors are commonly used for tracking of charged particles \cite{sauli,altunbas}. 

%%%%%%%%%%%%%%%%%%%%%%%%%%%%%%%%%%%%%%%%%%%%%%%%%%%%%%%%%%%%%
%%%%%%%%%%%%%%%%%%%%%%%%%%%%%%%%%%%%%%%%%%%%%%%%%%%%%%%%%%%%%

Upcoming experiments such as the Compressed Baryonic Matter (CBM) at the future Facility for Antiproton and Ion Research (FAIR) in Darmstadt, Germany \cite{cbm,fair,cbmweb} will use GEM detectors in the  Muon Chamber (MuCh) system, particularly in the first and second stations where the particle flux is extremely high ($\sim$\,1~MHz/cm$^2$) \cite{adak, chatterjee1}. According to the simulation study reported for the CBM MuCh detector, the particle hit rates at a beam kinetic energy of 12 AGeV (12 GeV per nucleon, where $A$ denotes the mass number of the incident nucleus) are expected to be about 290~kHz/cm$^2$, 48~kHz/cm$^2$, 20~kHz/cm$^2$, and 10~kHz/cm$^2$ for the first, second, third, and fourth stations, respectively \cite{sharma2023}. Under such harsh radiation condition, detectors needs to be operated stably with high efficiency, stable gain and good energy resolution over long period of time. Therefore, it is essential to study the detector behaviour under different operating conditions and to identify an ideal working region. The performance of a GEM detector vary with the variation in the applied voltage across the GEM foils. The electric field inside the GEM holes depends on this bias voltage, and even small variations can significantly affect the avalanche multiplication process \cite{bachmann,shah}. As a result, important detector parameters such as gain, energy resolution and count rate show strong dependence on the applied voltage~\cite{adak, mandal}. Understanding of these dependences are very important. In this work, a detailed voltage scan study is carried out using a $^{55}$Fe X-ray source. The analysis focuses on the voltage scan, where energy spectra have been recorded over a range of $\Delta V$ values. The change of the spectra with increasing bias voltage, including the shift in peak position, peak broadening provides valuable understanding of the gain characteristics and signal formation mechanism of the detector \cite{roy,sen}. At lower voltages, the detector shows weak signal response due to insufficient electron multiplication, resulting in broad spectra with poorly defined peaks. As the applied voltage increases, the electric field inside the GEM holes becomes stronger, leading to more avalanche multiplication and higher signal amplitude. Consequently, the spectral peaks become sharper and more clearly defined. At sufficient high voltages, both the 5.9~keV full energy peak and the Argon escape peak are clearly visible, showing improved detector performance~\cite{chatterjee2,shah}. Such systematic studies are essential to identify the most suitable operating region where the detector achieves a balance between high gain, good energy resolution and high efficiency. These characteristics are particularly important for the  use of GEM detectors in the high-rate experiments such as the CBM. In this paper, characteristic study of a single-mask triple-GEM detector carried out using a $^{55}$Fe X-ray source with an Ar/CO$_2$ (70/30) gas mixture is described. Effects after long-term studies are emphasised.

%%%%%%%%%%%%%%%%%%%%%%%%%%%%%%%%%%%%%%%%%%%%%%%%%%%%%%%%%%%%

\section{Detector and Experimental Setup}

The detector used in this study is a 10~$\times$~10~cm$^2$ single-mask triple-GEM detector with a 3-2-2-2 gap configuration. The drift gap is 3 mm, while the transfer and induction gaps are kept at 2 mm each as shown in Fig~\ref{fig1}. The detector is operated using an Ar/CO$_2$ gas mixture having a volume ratio of 70/30 at a flow rate of approximately 3.5-4~l/h. A strong $^{55}$Fe mono-energetic X-ray source having characteristic energy of  5.9~keV  is used for irradiation using a collimator made of G-10 material and having a hole of area $\sim$~50~mm$^2$. The full absorption of a 5.9~keV X-ray photon in the drift region filled with Ar/CO$_2$ gas mixture in 70/30 volume ratio, produces an average of approximately 212 primary electrons. The primary electrons generated in the drift region move towards the first GEM foil under the influence of the drift field. As they pass through successive GEM layers, they undergo multiplication at each stage, resulting in a cascade amplification process. The multiplied electrons after passing through the induction region are finally collected at the readout plane where they induce a measurable signal. 

The high voltage (HV) required for detector operation is supplied through a resistive divider chain (as shown in Fig~\ref{fig1}), which ensures proper distribution of voltage across the drift, transfer and induction gaps as well as across the GEM foils. 256-X and 256-Y strips are used to collect signals from X and Y. However, 4 sum-up boards (each board connect 128 strips, 2 boards for X and 2 boards for Y), provided by CERN are used to sum the strips and collect signals. The signals from 128 readout strips collected through one sum-up board, are processed using a charge-sensitive preamplifier (VV50-2), having a gain of 2~mV/fC and a shaping time of 300~ns~\cite{preamplifier}, which converts the collected charge into a voltage pulse. This signal is further shaped and analysed using a Multi-Channel Analyser (MCA) to obtain the energy spectra. The count rate is measured using a scaler module, which records the number of detected events over a given time interval. The details of the electronics set-up is given in Ref.\cite{s_chatterjee_charging_up_1,mandal1}. All measurements are carried out in the laboratory along with the monitoring and recording of ambient parameters such as temperature, pressure and relative humidity, which are known to change the gas gain and other performance of gas detectors \cite{mandal}.

\section{Results and Discussion}
%%%%%%%%%%%%%%%%%%%%%%%%%%%%%%%%%%%%%%%%%%%%%%%%%%%%%%%%%%%%%

\begin{figure}[htb]
\centering
\includegraphics[scale=0.5]{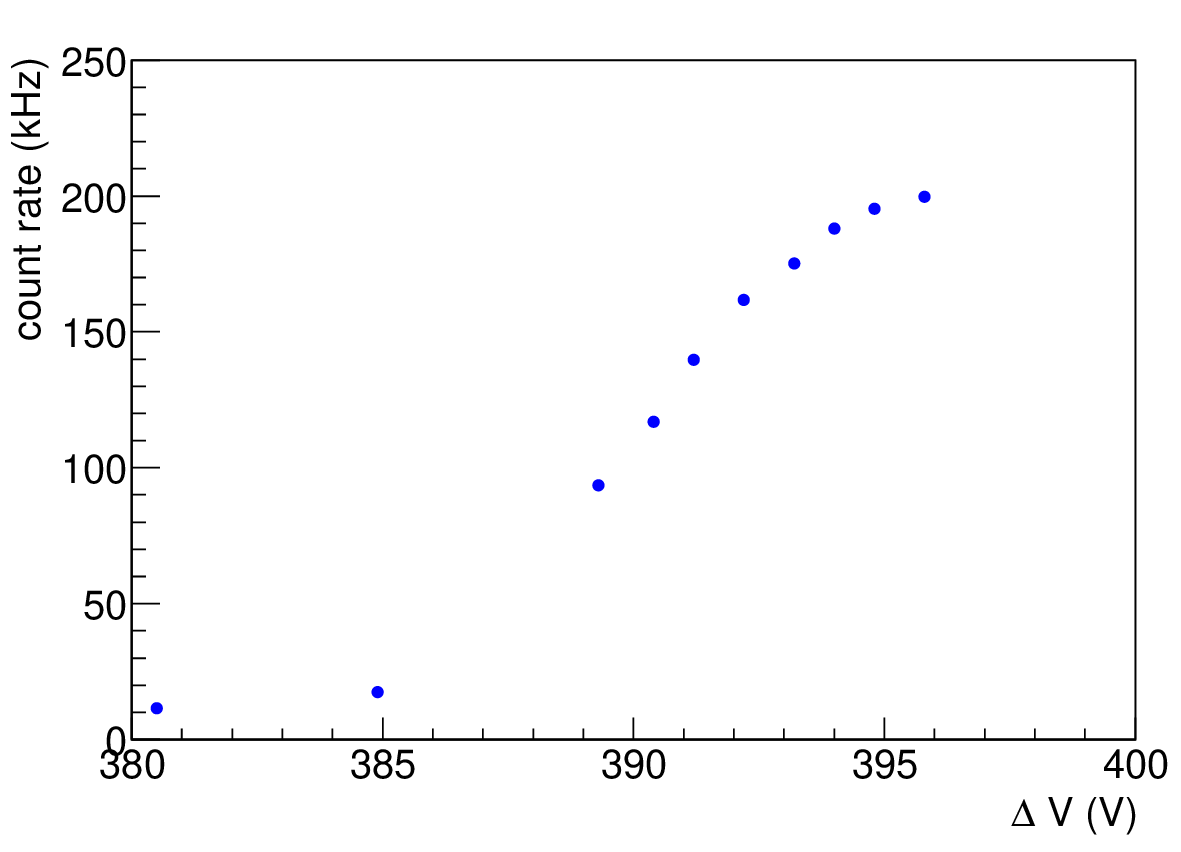}
\caption{\label{DeltaV_area} Count rate with the source as a function of the $\Delta V$ across each GEM foil. Error bars are smaller than the marker size.}
\end{figure}

%%%%%%%%%%%%%%%%%%%%%%%%%%%%%%%%%%%%%%%%%%%%%%%%%%%%%%%%%%%%%
%%%%%%%%%%%%%%%%%%%%%%%%%%%%%%%%%%%%%%%%%%%%%%%%%%%%%%%%%%%%%

\begin{figure}[htb]
\centering
\includegraphics[scale=0.5]{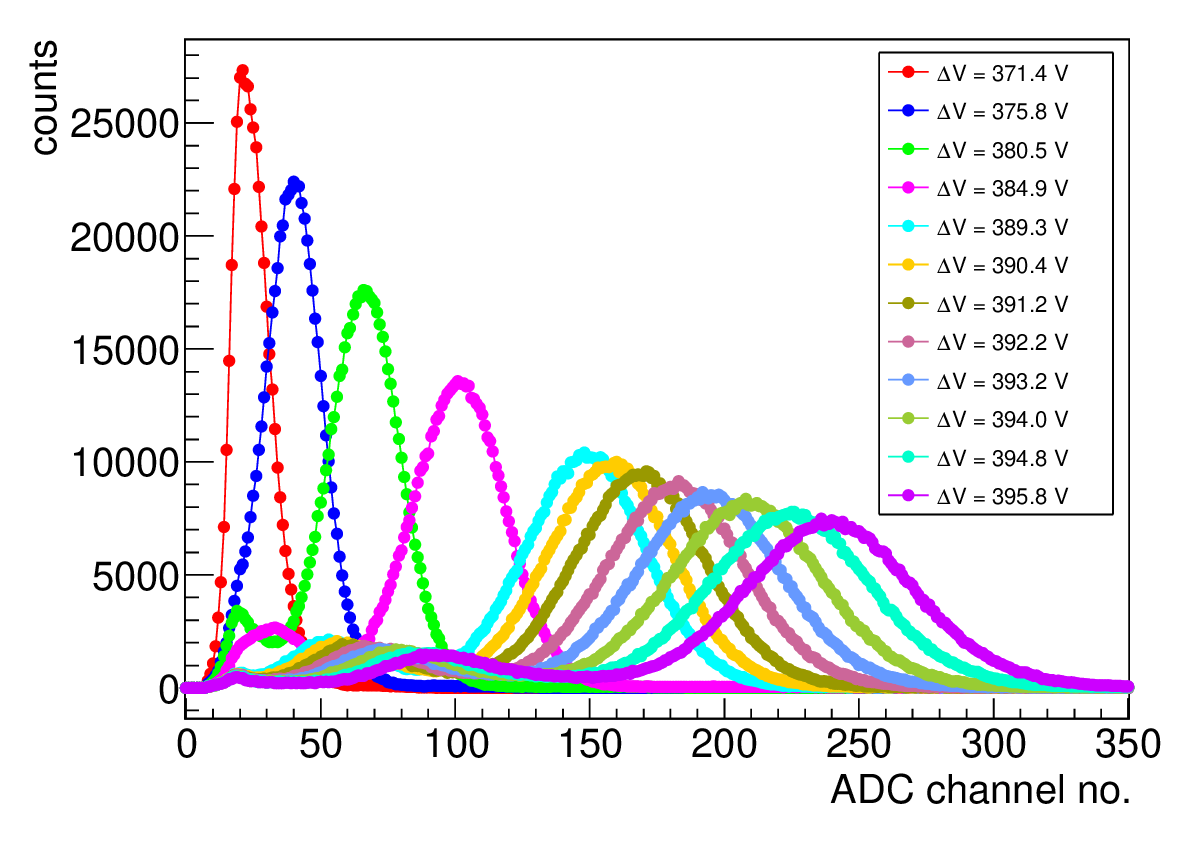}
\caption{\label{volt_scan_plot}Energy spectra of $^{55}$Fe X-ray at different  $\Delta V$ across each GEM foil.}
\end{figure}
%%%%%%%%%%%%%%%%%%%%%%%%%%%%%%%%%%%%%%%%%%%%%%%%%%%%%%%%%%%%%

%\subsection{Count Rate Behaviour}
In this work the count rate with X-ray source which is a figure of merit \cite{mandal1} for efficiency is studied as a function of the applied $\Delta V$ across each GEM foil. The count rate as a function of $\Delta V$ is shown in Fig~\ref{DeltaV_area}. The count rate increases rapidly with increasing $\Delta V$ across the GEM foil, due to the increase in avalanche multiplication inside the GEM holes. At low voltages, the weak electric field leads to insufficient signal generation and lower count rate. As the voltage increases, more events exceed the detection threshold, resulting in higher count rates. A saturation is observed in the higher voltage regime, where the count rate becomes nearly constant at $\sim$200~kHz. This HV region indicates maximum detection efficiency defining the optimal working range.

%\subsection{Voltage Scan Behaviour}

The energy spectra of $^{55}$Fe X-ray at different  $\Delta V$ are shown in Fig.~\ref{volt_scan_plot}. With increasing $\Delta V$ (from $\sim$371.4~V to $\sim$395.8~V), the 5.9~keV main peak shifts towards higher ADC channels and the mean value of the main peak also increases indicating an increase in the gas gain. As $\Delta V$ increases the ratio of the full width at half maxima (FWHM) and the mean value of the main peak also decreases indicating better energy resolution. The clear separation of the 5.9~keV main peak and $\sim$2.9~keV escape peaks at higher $\Delta V$ reflects that improvement in energy resolution.

%%%%%%%%%%%%%%%%%%%%%%%%%%%%%%%%%%%%%%%%%%%%%%%%%%%%%%%%%%%%%
\begin{figure}[htb]
\centering
\includegraphics[scale=0.5]{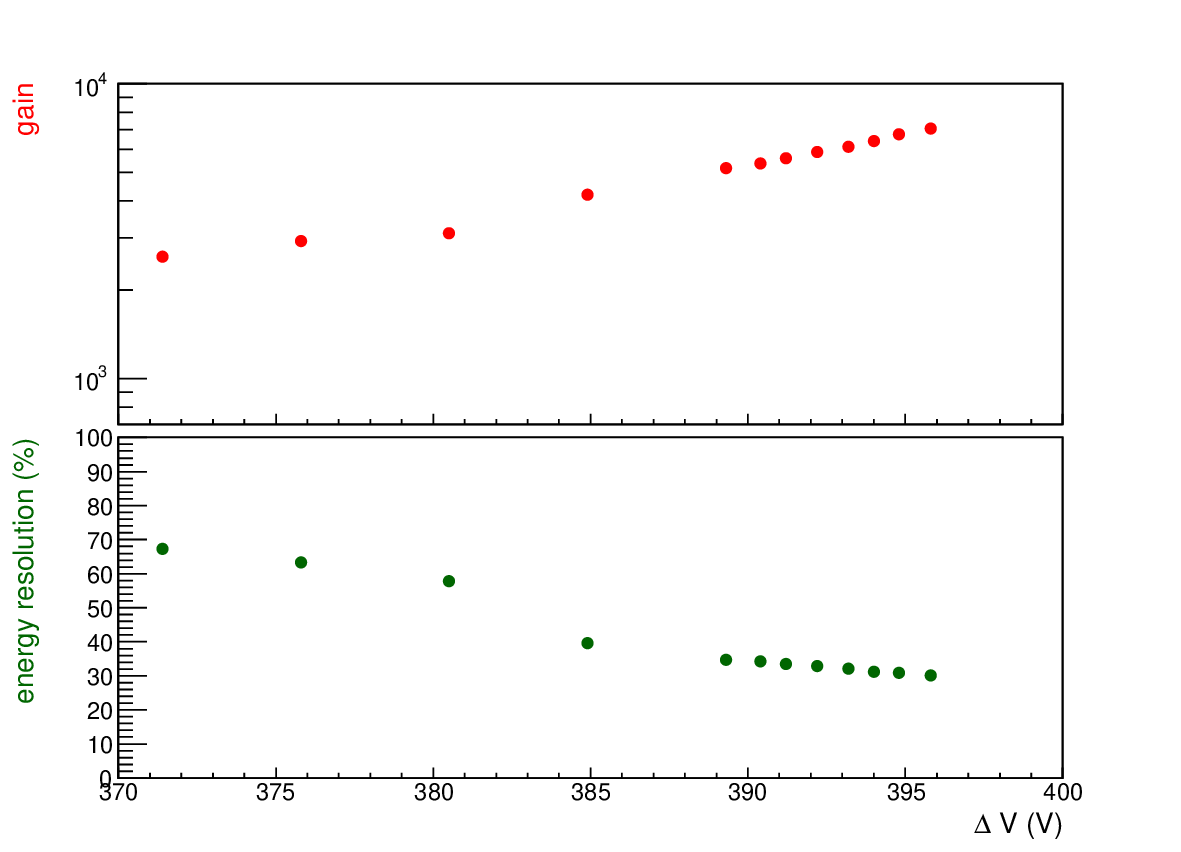}
\caption{\label{deltaV_gain_reso_log}Gain and energy resolution as a function of the $\Delta$V. Error bars are smaller than the marker size. }
\end{figure}

%%%%%%%%%%%%%%%%%%%%%%%%%%%%%%%%%%%%%%%%%%%%%%%%%%%%%%%%%%%%
%%%%%%%%%%%%%%%%%%%%%%%%%%%%%%%%%%%%%%%%%%%%%%%%%%%%%%%%%%%%%
\begin{figure}[htb]
\centering
\includegraphics[scale=0.44]{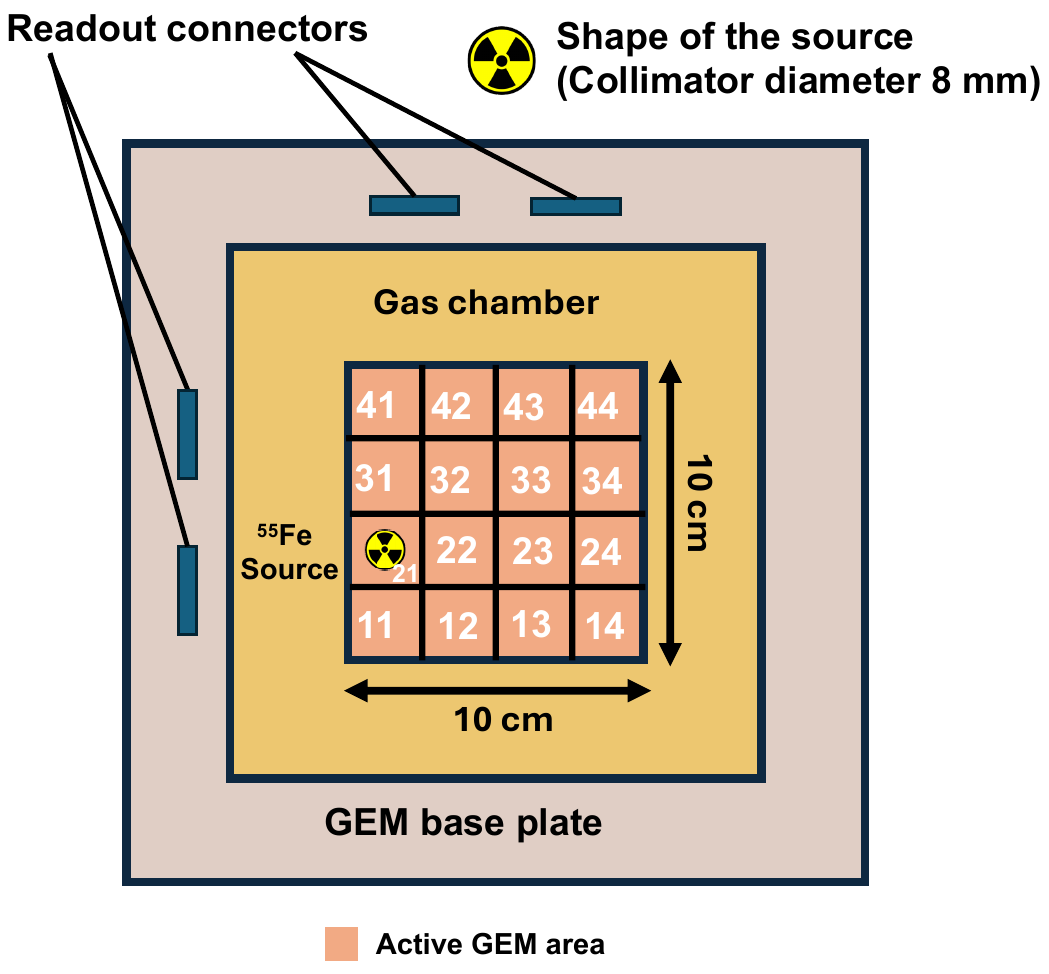}
\caption{\label{segment} Schematic of the segmentation of the triple-GEM detector.}
\end{figure}
%%%%%%%%%%%%%%%%%%%%%%%%%%%%%%%%%%%%%%%%%%%%%%%%%%%%%%%%%%%%%

%%%%%%%%%%%%%%%%%%%%%%%%%%%%%%%%%%%%%%%%%%%%%%%%%%%%%%%%%%%
\begin{figure}[htb]
\centering
\includegraphics[scale=0.5]{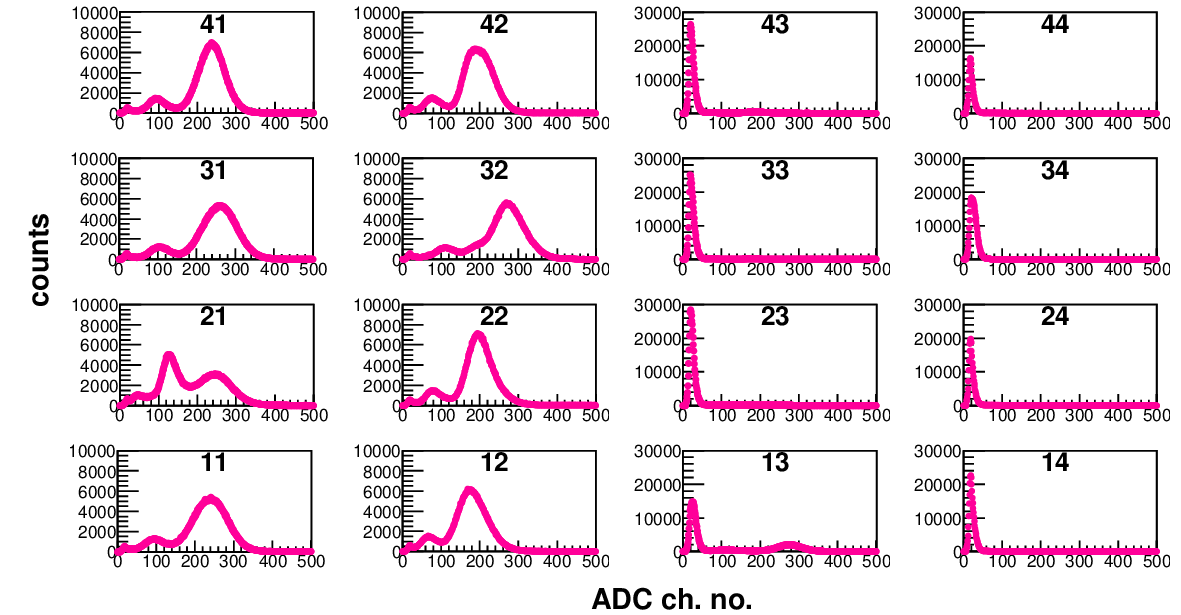}
 	\caption{\label{pulse} $^{55}$Fe spectra at an applied $\Delta$V of $\sim$396~V from 16 different places over the detector.}
\end{figure}

%%%%%%%%%%%%%%%%%%%%%%%%%%%%%%%%%%%%%%%%%%%%%%%%%%%%%%%%%%%%

%\subsection{Gain and Energy Resolution}
The gain and energy resolution (as defined in Ref.~\cite{mandal1} and \cite{mandal_long_term_stability_1}) with applied $\Delta V$ across the GEMs are shown in Fig.~\ref{deltaV_gain_reso_log}. The gain increases exponentially from $\sim$2600 to $\sim$7050 with increasing $\Delta V$ from $\sim$371~V to $\sim$396~V, due to the avalanche multiplication process in the GEM holes.  The energy resolution improves from approximately 67\% to 30\% with increasing $\Delta V$.

%\subsection{Positional Uniformity Study}
One of the main emphasis of this study is to check the aftereffect of the long-term irradiation. In this direction, the entire area of the GEM detector is divided into a 4$\times$4 imaginary grid, with positions labeled 11, 12, 13, 14; 21, 22, 23, 24; 31, 32, 33, 34 and 41, 42, 43, 44 as shown in Fig.~\ref{segment}. It is to be mentioned here that positions 21 and 12 are exposed to long-term irradiation during stability studies for $\sim$95~days \cite{mandal_long_term_stability_1} and $\sim$20~days \cite{mandal} respectively, at an average rate of $\sim$200~kHz. In this position uniformity study, data are collected using a single sum-up board, and the $^{55}$Fe source is moved step by step over each of the 16 segments. During the entire uniformity study the GEM is operated at a $\Delta$V of $\sim$396~V across each foil. The X-ray spectra placing the $^{55}$Fe source at different places are shown in Fig.~\ref{pulse}. It is observed from Fig.~\ref{pulse} that reasonable X-ray spectra are obtained for only 8 segments (Left part) because in this setup, 128 readout strips (X strips) are connected to a single sum-up board which is connected to the pre-amplifier for signal collection. The remaining 3 sum-up [1 for X strips (Right part) and 2 for Y strips] boards are terminated with 50~$\Omega$ resistors. As these channels are not connected to the readout electronics (pre-amplifier from which the signals are collected), no measurable signals are expected from them, and consequently, no spectra are obtained from this section of the detector. In these spectra (only from left 8 segments), the 5.9 keV full energy peak and the Argon escape peak are clearly visible. The spectrum corresponding to position 21 is somewhat distorted compared with those obtained from the other measured regions. Position 21 had previously been subjected to approximately 95 days of continuous high-rate irradiation. It is to be mentioned here that this particular uniformity measurement is carried out immediately after stopping the continuous long-term test of 95~days keeping X-ray source at 21 position. %Together with the lower gain and source count rate observed at this position, the spectral distortion indicates a local change in detector response following prolonged irradiation. Such a change may arise from irradiation-induced ageing and/or modifications of the local charge-amplification conditions, however, the present measurements do not allow the microscopic origin of the effect to be identified unambiguously.

%%%%%%%%%%%%%%%%%%%%%%%%%%%%%%%%%%%%%%%%%%%%%%%%%%%%%%%%%%%%

\begin{figure}[htb]
\centering
\includegraphics[scale=0.5]{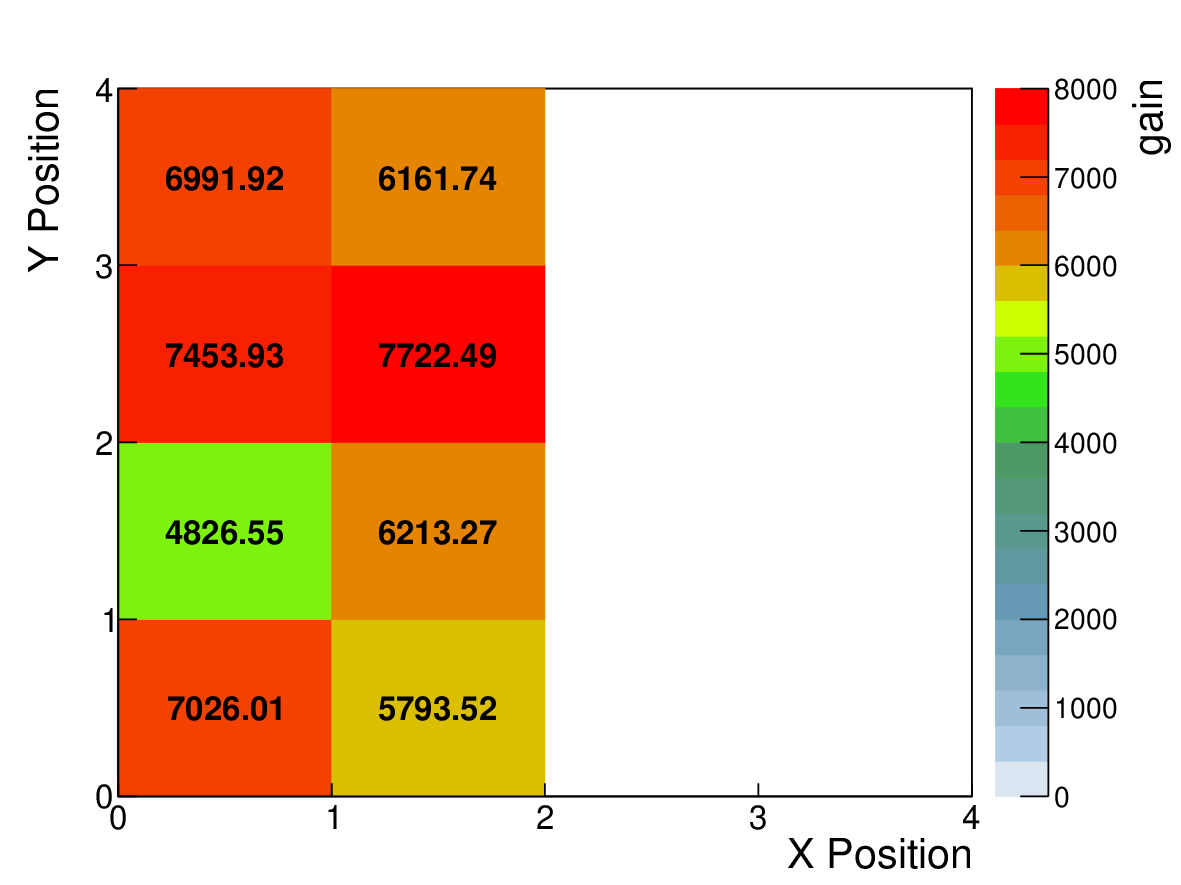}
	\caption{\label{gain} Gain at 8 different positions on the connected region of the detector.}
\end{figure}

%%%%%%%%%%%%%%%%%%%%%%%%%%%%%%%%%%%%%%%%%%%%%%%%%%%%%%%%%%%%
%%%%%%%%%%%%%%%%%%%%%%%%%%%%%%%%%%%%%%%%%%%%%%%%%%%%%%%%%%%%
\begin{figure}[htb]
\centering
\includegraphics[scale=0.5]{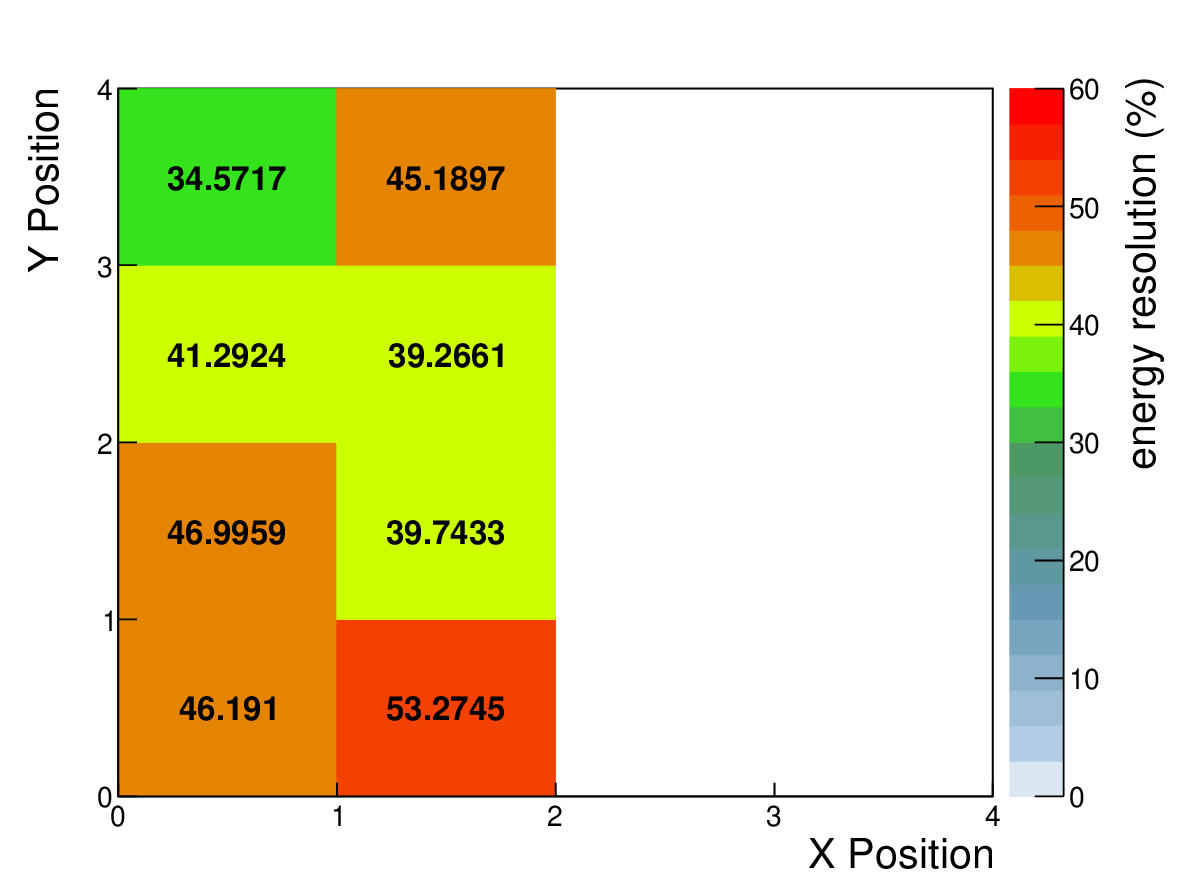}
	\caption{\label{nor_gain} Energy resolution at 8 different positions on the connected region of the detector.}
\end{figure}

%%%%%%%%%%%%%%%%%%%%%%%%%%%%%%%%%%%%%%%%%%%%%%%%%%%%%%%%%%%%
%%%%%%%%%%%%%%%%%%%%%%%%%%%%%%%%%%%%%%%%%%%%%%%%%%%%%%%%%%%%
\begin{figure}[htb]
\centering
\includegraphics[scale=0.5]{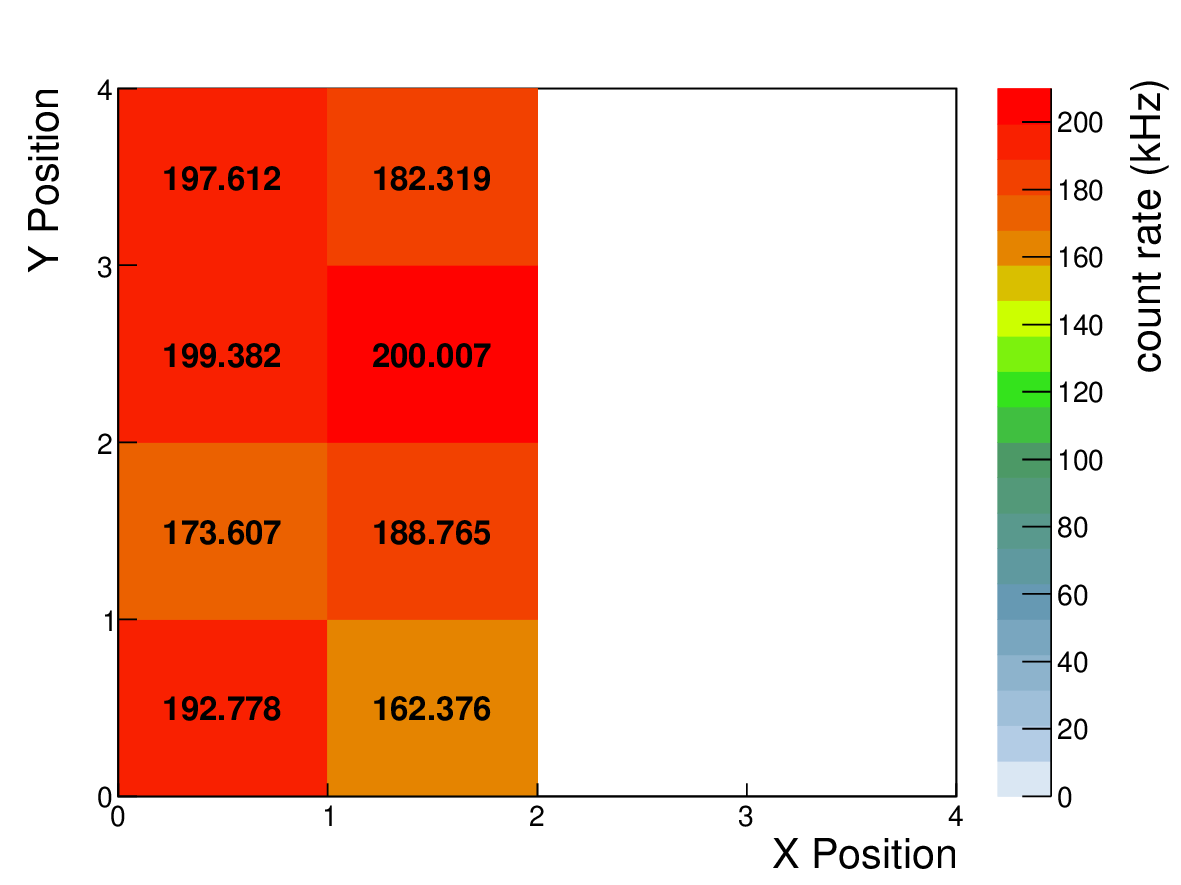}
	\caption{\label{pulse_2} Count rate with source at 8 different positions on the connected region of the detector.}
\end{figure}
%%%%%%%%%%%%%%%%%%%%%%%%%%%%%%%%%%%%%%%%%%%%%%%%%%%%%%%%%%%%

%%%%%%%%%%%%%%%%%%%%%%%%%%%%%%%%%%%%%%%%%%%%%%%%%%%%%%%%%%%%
\begin{figure}[htb]
\centering
\includegraphics[scale=0.5]{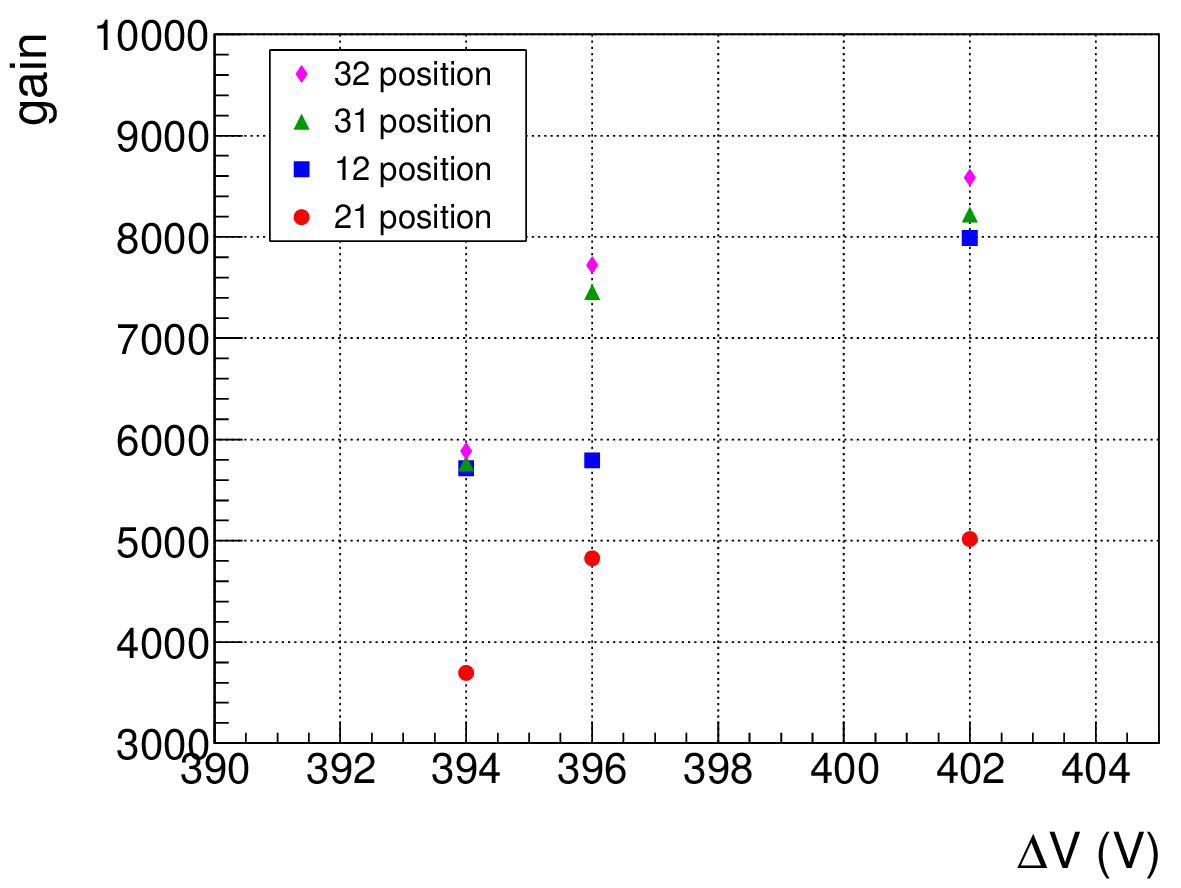}
	\caption{\label{aged} Gain as a function of $\Delta$V at 4 positions after the long-term stability test.}
\end{figure}

%%%%%%%%%%%%%%%%%%%%%%%%%%%%%%%%%%%%%%%%%%%%%%%%%%%%%%%%%%%%

The gain measured at different positions on the detector (left part) at $\Delta$V of 396~V is shown in Fig.~\ref{gain}. The gain is determined by fitting the full energy peak of the $^{55}$Fe spectra with a Gaussian function, as described in Ref.~\cite{mandal}. It is observed that, positions 21 and 12 show gain values of $\sim$4800 and $\sim$5800 respectively which are comparatively lower than the other positions. %The lower gain indicates reduced charge amplification, which may result from local variations in the electric field or degradation of the multiplication process due to prolonged irradiation.

The energy resolution at different positions of the detector is shown in Fig.~\ref{nor_gain}. It is observed that, positions 21 and 12 show energy resolution values of $\sim$47\% and $\sim$53\% respectively, which are relatively higher than the other positions. %This degradation in resolution can be due to prolonged irradiation to high rates. 

The count rate with the radioactive source measured at different positions is shown in Fig.~\ref{pulse_2}. Positions 21 and 12 show comparatively lower count rates of approximately 173~kHz and 162~kHz, respectively, whereas values close to 200~kHz are observed in the best-performing regions. Since the  count rate is used here only as a figure of merit for the detector response, these measurements should not be regarded as a direct determination of the absolute detection efficiency. Nevertheless, the combined observation of lower gain, poorer energy resolution and reduced  count rate at positions 21 and 12 demonstrates a local effect of prolonged irradiation on the performance of the detector. %A dedicated efficiency measurement with an independent reference detector would be required to determine whether these irradiated regions still satisfy the efficiency requirement of a specific experiment.

%The comparatively lower gain and count rate observed at positions 21 and 12 are correlated with their prolonged irradiation histories and are therefore interpreted as local irradiation-induced ageing or degradation. 

It should be noted here that ageing in gaseous detectors depends not simply on elapsed irradiation time but on the accumulated charge and operating conditions. In the long-term study of the same single-mask triple-GEM chamber, no continuous degradation was observed up to an accumulated charge density of 1.35~mC~mm$^{-2}$~\cite{mandal} at 12 position and $\sim$8~mC~mm$^{-2}$~\cite{mandal_long_term_stability_1} at 21 position. The same study also showed that changes in the divider current can modify the effective voltage across the GEM foils and consequently the measured gain. Therefore, although the present local reduction at positions 21 and 12 is consistent with irradiation-induced effect, the available measurements do not allow the underlying mechanism to be attributed uniquely to classical polymer-deposition ageing.

Ageing studies of gaseous detectors show that accumulated charge, rather than irradiation time alone, is the more relevant parameter for comparing radiation exposure, while the observed ageing rate can depend strongly on dose rate, irradiated area, gas flow, detector materials and trace contaminants~\cite{sauli_book}. Micro-pattern gaseous  detectors, and GEMs in particular, are generally found to be comparatively radiation tolerant. Accelerated ageing tests of GEM detectors operated with Ar/CO$_2$ have reported no significant degradation up to accumulated charge densities of the order of 1~C~cm$^{-2}$~\cite{sauli_book}. Consequently, a direct quantitative comparison based only on the 95~day and 20~day irradiation periods of the present study would not be appropriate. The local changes observed here should instead be considered a combined effect of the accumulated charge and the long-term operational history of this particular chamber~\cite{mandal,sauli_book}.

To check whether after the long-term irradiation the chamber is still sensitive or not in some of the zones including the irradiated zones the gain is measured increasing the voltage. Fig.~\ref{aged} shows the gain at four different positions namely 32, 31, 12 and 21, measured at three different voltage settings, after the long-term stability test. It is mentioned earlier that the long-term irradiation are made at 12 and 21 positions. It is observed from Fig.~\ref{aged} that, although the value of gain at 12 and 21 are comparatively a bit lower, it increases with the increase of voltage in those positions as well like other positions which are not irradiated for long. It means those places of GEM chambers where the radiations are exposed for long time are still quite sensitive to applied voltage.

\section{Summary}
A details performance study of a single-mask triple-GEM detector is carried out using a $^{55}$Fe X-ray source with an Ar/CO$_2$ gas mixture in 70/30 volume ratio. The detector characteristics are investigated as a function of the applied voltage, focusing on key parameters such as gain, energy resolution and count rate. The gain is observed to increase exponentially with the applied voltage due to increased avalanche multiplication within the GEM holes. On the other hand the energy resolution improves with voltage. The count rate increases with voltage and reaches a saturation region, indicating that the detector achieves its highest efficiency within a specific voltage range. The voltage scan study clearly shows that the main peak move to higher ADC channel and the separation of the full energy peak and Argon escape peak at higher voltages becomes more prominent. This indicate that the more effective charge multiplication and signal formation of the detector is obtained at higher voltages. 

The performance over the surface study shows that well defined spectra are obtained only from that part (Left part), from which the read-out strips are connected to preamplifier. When the source is placed to the zones on the other part (Right part) of the chamber and the sum-up board of the other part is connected to the preamplifier no spectra are obtained as expected. Most of the investigated regions of the detector (Left side) maintain stable gain, energy resolution and  count rate. In contrast, positions 21 and 12, which are subjected to prolonged high-rate irradiation, exhibit little different behaviour. The gain at these positions nevertheless continues to increase with applied voltage, showing that the irradiated regions remain responsive and that charge multiplication is still present and increases with the increase of applied voltage. The observed local behaviour of those two positions, therefore, highlights the importance of monitoring gain, energy resolution, efficiency and accumulated charge during prolonged irradiation tests. Further dedicated efficiency and stability studies are required to quantify the remaining performance changes of the irradiated regions.
%%%%%%%%%%%%%%%%%%%%%%%%%%%%%%%%%%%%%%%%%%%%%%%%%%%%%%%%%%%%

\section{Acknowledgements}	
The authors would like to thank the RD51 collaboration (presently DRD1) for the support in building and initial testing of the chamber at CERN. The authors would also like to thank Dr. Arindam Sen of Jagiellonian University, Poland, Dr. Somen Gope of Warsaw University of Technology, Poland and Dr. Sayak Chatterjee of  University of Massachusetts, Amherst, USA for valuable discussions during the work. The authors acknowledges Mr. Subrata Das for helping in building the collimators and other mechanical systems used in this study. This work is partially supported by the research grant SR/MF/PS-01/2014-BI from DST, Govt. of India, and the research grant of the CBM-MuCh project from BI-IFCC, DST, Govt. of India. S. Mandal acknowledges his UGC-NET fellowship (NTA Ref. No.: 221610099585, Dt. 29.11.2022) for the support. S. Biswas acknowledges the support of the DST-SERB Ramanujan Fellowship (D.O.No. SR/S2/RJN-02/2012).
%%%%%%%%%%%%%%%%%%%%%%%%%%%%%%%%%%%%%%%%%%%%%%%%%

%%%%%%%%%%%%%%%%%%%%%%%%%%%%%%%%%%%%%%%%%%%%%%%%%

\end{document}